# Vertical GaN Devices: Process and Reliability

Shuzhen You[a,*], Karen Geens[a], Matteo Borga[a], Hu Liang[a], Herwig Hahn[b], Dirk Fahle[b], Michael Heuken[b], Kalparupa Mukherjee[c], Carlo De Santi[c], Matteo Meneghini[c], Enrico Zanoni[c], Martin Berg[d], Peter Ramvall[d], Ashutosh Kumar[d], Mikael T. Björk[e], B. Jonas Ohlsson[e], Stefaan Decoutere[a]

[a] *imec, Leuven, Belgium*
[b] *AIXTRON SE, D-52134 Herzogenrath, Germany*
[c] *Department of Information Engineering, University of Padua, 35131 Padova, Italy*
[d] *RISE Research Institutes of Sweden, Scheelevägen 17, S-223 70, Lund, Sweden*
[e] *Hexagem AB, Ideon Alfa 3, Scheelevägen 15, S-223 63 Lund, Sweden*

**Abstract**

This paper reviews recent progress and key challenges in process and reliability for high-performance vertical GaN transistors and diodes, focusing on the 200 mm CMOS-compatible technology. We particularly demonstrated the potential of using 200 mm diameter CTE matched substrates for vertical power transistors, and gate module optimizations for device robustness. An alternative technology path based on coalescence epitaxy of GaN-on-Silicon is also introduced, which could enable thick drift layers of very low dislocation density.

## 1. Introduction

GaN power devices have been widely recognized as promising candidates for next generation power electronics. High voltage blocking capability is key for efficiency gains in GaN power electronics. Currently, both lateral GaN HEMT and vertical GaN MOSFET are considered for achieving high breakdown voltage beyond 650V. When aiming at high voltage, lateral GaN devices are limited by the area occupancy and surface trap related reliability concerns. Contrary to the lateral GaN devices, vertical GaN devices do not require enlarged chip size for increased breakdown voltage. The breakdown voltage in vertical GaN devices can be increased by increasing the thickness of the drift region while keeping a compact chip size. Moving the peak electric field from the surface into the GaN bulk potentially minimizes the surface trapping effect and dynamic Ron in the vertical devices [1][2].

Today, GaN technology has presented transistors achieving breakdown voltages up to 1700 V and diodes with breakdown voltage up to 4000 V [3][4][5][6][7][8][9]. However, they have mostly been demonstrated only in research labs, and require special techniques/treatments that might not be suited for large scale fabrication. This paper reviews the progress and addresses key challenges in process and reliability, focusing on 200 mm CMOS-compatible technology.

We present 2 possible large diameter substrates for vertical GaN fabrication, which pave the way for industrial fabrication of this type of devices. Furthermore, the optimization of the doping level in the GaN device stack is discussed and finally the trench gate processing and the reliability of the fabricated MOSFETs are discussed.

## 2. Large diameter substrates for vertical GaN fabrication

*2.1. Engineered CTE matched substrates.*

State-of-the-art vertical GaN devices are fabricated on bulk GaN substrates, thanks to the high quality of the substrates in terms of low dislocation density and low impurity concentrations [10]. However, they are prohibitively expensive, and only small area substrates are available. Therefore, an industrially scaled process on bulk GaN substrates is not foreseen in near future. On the other end of the spectrum, several research groups have demonstrated (quasi-)vertical diodes and transistors on GaN-on-Si substrates [11][12][13][14][15][16][17]. However, these devices have a limited drift layer thickness or a limited wafer diameter due to the large mismatch in coefficient of thermal expansion (CTE) between Si and GaN. QST® substrates, with a CTE matched poly-AlN core, can overcome these limitations and offer a breakthrough in growth of thick GaN epitaxial layer.

___
\* Corresponding author. Shuzhen.you@imec.be
Tel: +32 (16) 281104.

Qromis Substrate Technology (QST®) has been a pioneer in developing engineered substrates with SEMI standard thickness. The QST® substrate, depicted schematically in Fig. **1**(a), includes a CTE matched poly-AlN core; engineered layers; buried oxide (BOX) and monocrystalline Si(111) serving as the nucleation layer for the metal-organic chemical vapour deposition (MOCVD) [18].

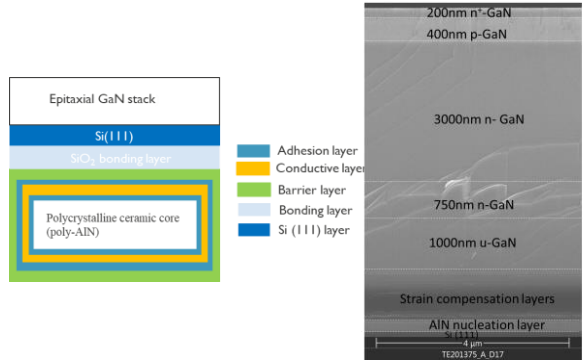

Fig. 1. (a) Schematic image of QST® engineered substrates with poly-AlN core (with epitaxial GaN stack on top) , (b) Cross-sectional SEM image a vertical device stack with a total GaN thickness of 5.4 µm on 200mm engineered substrates.

The potential of engineered QST® substrates for growing thick epitaxial layer and large wafer diameter, has been validated by several groups, by growing >10 µm, crack-free, device-quality epitaxial layers on 150mm substrates and 6 µm stack on 200mm substrates [19][20][21][22]. By introducing a dislocation reduction layer on these substrates, a low dislocation density in the order of $10^7/cm^2$ can be achieved[23][24]. Furthermore, these substrates have been used for fabrication of enhancement mode p-GaN gate HEMT devices, both for discrete components and monolithically integrated circuits, with (Al)GaN layers grown by means of MOCVD. Employing material grown in an AIX G5+ C Planetary Reactor® from AIXTRON, Germany, lateral devices including ICs were fabricated using imec's advanced 200mm CMOS compatible Si pilot line [25][26][27][28]. QST® engineered substrates will be commercially available from both Shin-Etsu and Qromis for industry players [29][30] and can be scaled up in diameter to 300mm. By 2025 the price is expected to approach the one of SOI wafers, which paves the way for industrial fabrication of GaN power devices beyond 1200 V.

With respect to vertical GaN device fabrication, we demonstrated a crack-free MOCVD-grown 5.4 µm-thick GaN stack grown also in an AIX G5+ C Planetary Reactor®. The stack, shown in Fig. 1(b), consists of a 200 nm AlN nucleation layer, a strain-compensating buffer, a 1 µm unintentionally doped GaN layer, a 750 nm $n^+$-GaN bottom contact, a 3 µm-thick $n^-$ GaN drift layer, a 400 nm Mg-doped p-GaN channel layer and a 200 nm $n^+$-GaN top contact layer. Careful engineering of the strain-compensating buffer layer is required to obtain an optimal crystal quality of the GaN-based stack. In this work we present two generations of vertical device stacks on engineered substrates. The difference in crystal quality is evident from XRD data as listed in Table 1. The XRD FWHM value was improved from 627 to 406 arcsec for the GaN <102> peak, indicating we were able to reduce the edge type dislocation density by 60% to an estimated amount of $4\text{-}5\times10^8$ cm$^{-2}$ [31]. The Si doping level from the generation 1 to 2, was reduced from $4\times10^{16}$ cm$^{-3}$ to $2\times10^{16}$ cm$^{-3}$ (Table 1).

|  | GaN<102> FWHM (arcsec) | GaN<002> FWHM (arcsec) | $N_{D,n\text{-drift}}$ (Si/cm$^3$) |
|---|---|---|---|
| Gen 1 | 627 | 260 | $4\times10^{16}$ |
| Gen 2 | 406 | 227 | $2\times10^{16}$ |

Table 1. XRD FWHM of GaN <102> and GaN <002>, comparing generation 1 & 2 epitaxy on engineered substrates, for a stack with a 3 µm thick drift layer, using a different Si doping level in the $n^-$-drift layer.

The blocking voltage capability of these two generation stacks were evaluated by quasi-vertical NP diode test structures. In comparison of generation 1, generation 2 resulted in an increased breakdown voltage from 440 V to 680 V, as shown in Fig. 2(a). Benchmarked against the results from the devices on GaN-on-Si and GaN-on-GaN substrates, shown in Fig. 2(b), the breakdown voltage of the diodes in this work follows the projection of breakdown voltage versus drift layer thickness, validating the GaN on engineered substrates as a good choice for vertical device fabrication.

*2.2. Coalesced nanowire GaN on Si substrates*

GaN-on-Si using nanowire technology is another potential route towards thick, planar GaN drift layers with low dislocation density. The method is scalable up to 200 mm wafers, at lower cost than standard technology due to the very thin buffer and coalesced layers needed, enabling 1200 V devices at a competitive price.

In this technology, key challenges are the pre-processing of growth templates for nanowire selective area epitaxy and coalescence of the GaN nanowires[32][33][34]. Nanowires are nucleated from patterned holes in a $Si_3N_4$ mask. Having small enough holes to effectively filter dislocations from the substrate is one of the critical elements to achieve low dislocation densities in the subsequent layers. A diffraction-based lithography method is developed for cost efficient scaling, as a full wafer can be exposed at once. Presently, holes with 110 nm diameter and a hole-to-hole coefficient of variance below 2.7% can be produced, as shown in Fig. 3(a). For coalescence on one quarter 2" GaN-on-Si template wafers, the layers show more than 90 % surface coalescence coverage, and a smooth surface of 0.35 nm RMS roughness on a 2-µm-thick drift layer on top of coalesced material and a threading dislocation density of $6\times10^8$ cm$^{-2}$, as shown in Fig. 3(b). These results lay the foundation for further improvements to nanowire coalescence, but also a platform for thick drift layer growth.

## 3. Doping in the vertical device stack

### 3.1. Doping optimization of the n- drift layer

Having a low background concentration in the drift layer is key for well-performing devices in the 1200 V-class and beyond. To reach the 1200 V breakdown voltage target, doping levels need to be below $1.5\times10^{16}$ cm$^{-3}$, ideally even at $1\times10^{16}$ cm$^{-3}$ [35][36][37]. For reliable operation, the n-type donor (here: Si) needs to be roughly three times as large as the compensating C acceptor, i.e. for a net doping level $n = N_d - N_a = 1\times10^{16}$ cm$^{-3}$, the C level needs to be in the order of $5\times10^{15}$ cm$^{-3}$. A SIMS analysis of the stack of generation 1, depicted in Fig. 4 provides details on the chemical concentration of the individual layers. The n$^+$-GaN layers are doped with $5.5\times10^{18}$ Si/cc. The p-GaN layer has a $1.2\times10^{19}$ Mg/cc doping level, with a net p-doping of $2\times10^{18}$ cm$^{-3}$. The drift layer has a Si doping level of $4\times10^{16}$ cm$^{-3}$ with a C background level of $1.5\times10^{16}$ cm$^{-3}$. By tuning the growth parameters, a good suppression of background carbon below $1\times10^{16}$ cm$^{-3}$ is achieved, allowing the reduction of the Si doping level to $2\times10^{16}$ cm$^{-3}$ in the stack of generation 2, this result in a boost of the breakdown voltage of diode, as depicted in Fig. 2(a).

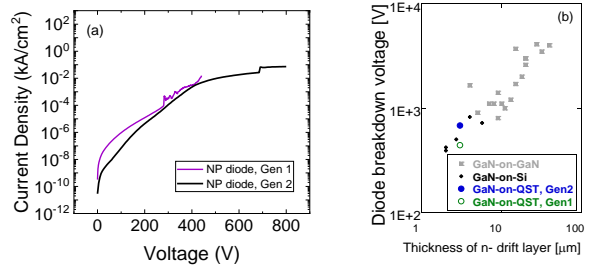

Fig. 2 (a) Diode leakage curves comparing generation 1 and generation 2 epitaxy on 200 mm engineered substrates (b) Benchmark the breakdown voltage of diodes versus the drift layer thickness on various substrates.

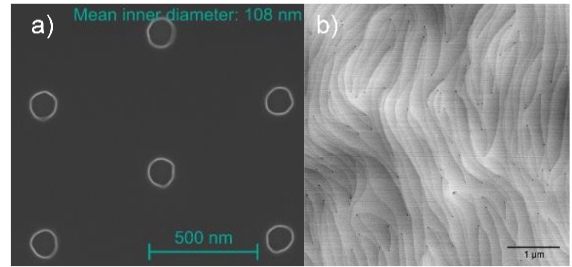

Fig. 3 (a) < 110 nm holes with a pitch of 600 nm in 60-nm-thick SiNx growth mask layer down to the GaN-on-Si substrate. (b) AFM on a 2-µm-thick drift layer on top of coalesced material for the same substrate. A dislocation density of $6\times10^8$ cm$^{-2}$ is estimated.

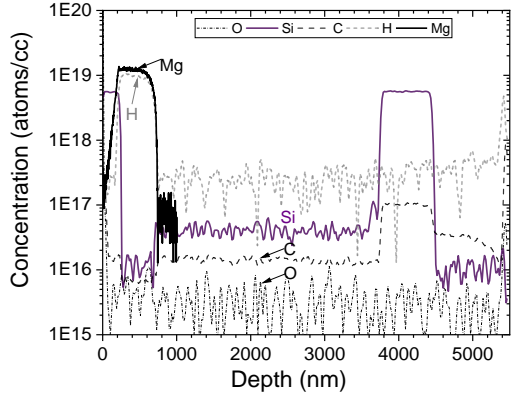

Fig. 4 SIMS profile of a stack of generation 1 with Si and C level of 4.0 and $1.5\times10^{16}$ cm$^{-3}$ respectively.

### 3.2. p-GaN channel doping and activation

While designing the p-GaN layer, a good blocking capability of the reverse biased PN-diode and a low channel resistance at the p-GaN/gate trench interface must be combined. The channel contribution to the device on-resistance, depends on both the channel length (which is defined by the p-GaN layer thickness) and on the inversion channel mobility (which increases with the p-doping). Fig. 5 depicts the correlation between $J_{D,SAT}$ and $R_{ON}$, showing reduced Ron and increased $I_{D,SAT}$ by reduction of p-GaN channel thickness and p-GaN Mg doping.

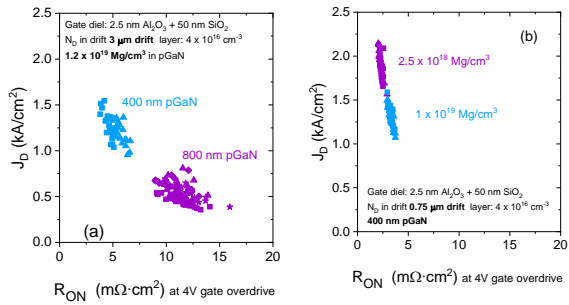

Fig. 5 $J_{D,SAT}$ versus $R_{ON}$ correlation plots comparing (a) devices on engineered substrates with a different p-GaN channel length and (b) devices on Si substrates, with a different Mg doping level in the p-GaN layer.

The output and transfer characteristic for a typical device with an 800 nm thick p-GaN layer is depicted in Fig. 6. For a channel length of 800 nm a current density of 0.35 kA/cm$^2$ is reached with a reasonably low $R_{ON}$ of 10 mΩ·cm$^2$.

On the other hand, the doping of the p-GaN layer can strongly impact the breakdown voltage. The p-GaN doping must be sufficiently high to avoid punch-through, and sufficiently low to reduce the peak electric field for device robustness [38][39].

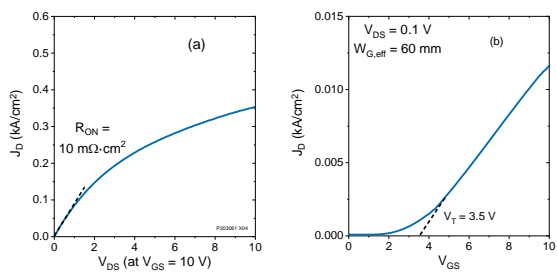

Fig. 6 Typical (a) output and (b) transfer characteristic of trench gate MOSFET device, with an 800 nm p-GaN channel processed on engineered substrates, with a poly-AlN core, using generation 1 GaN epitaxy.

## 4. Gate processing and device reliability

The gate dielectric in a (semi-) vertical trench MOSFET, shown in Fig. 7(a), must be chosen to achieve a low channel resistance and a high breakdown of the gate dielectric. Thick gate dielectrics are desired for robustness of the gate and higher transistor off-state breakdown voltage. On the other hand, thick dielectrics degrade the drive current and $R_{on}$. In our devices, a bilayer of a 2.5 nm Al$_2$O$_3$ and a 50 nm SiO$_2$ is used as gate dielectric. As shown in Fig. 7 (b), increasing the thickness of SiO$_2$ results in lower drive current and higher $R_{on}$.

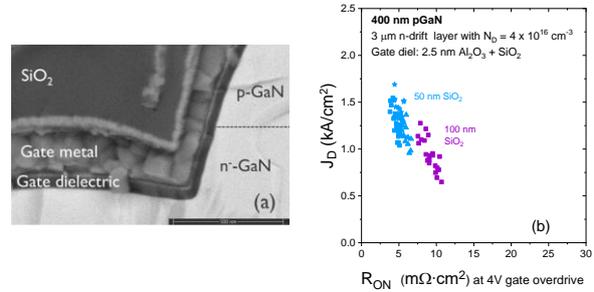

Fig. 7(a) Schematic cross-section of semi-vertical GaN MOSFET. (b) $J_{D,SAT}$ versus $R_{ON}$ correlation plots comparing devices on engineered substrates with different SiO$_2$ dielectric thickness in the gate trench region.

Special attention is paid to the processing of the gate trench, as a preferential failure occurs at the corners of the gate fingers, as was indicated by the drain step stress test until breakdown, coupled with electroluminescence (EL) studies[40][41], shown in Fig. 8. Such failures were identified to have been caused by an electrical breakdown of the gate isolation at the bottom edges of the trench [42] and was correlated with the presence of several abrupt steps of the gate trench sidewall. Therefore, an optimized ALE processing and wet cleaning sequence was implemented to achieve a smooth gate trench sidewall, shown in Fig. 7(a).

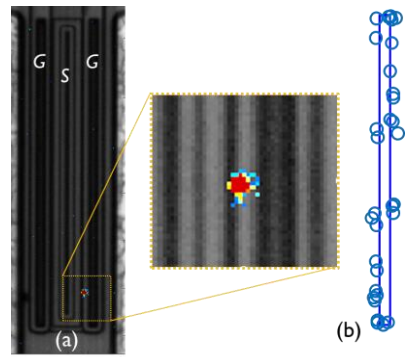

Fig. 8 Breakdown spot observed in $V_{DS}$ step stress test with electroluminescence imaging (a), and EL spot locations over both gate fingers (b).

## 5. Conclusions

In this review paper, we have proposed two promising substrate choices to enable industrial fabrication of vertical GaN devices on large diameter substrates. In a first part, 5.4 µm-thick high quality MOCVD-grown GaN epitaxial layers on 200 mm engineered substrates, with a diode blocking capability of 680 V were demonstrated. A second substrate choice, being coalesced GaN nanowires on Si, is scalable up to 200 mm and has potential to demonstrate GaN growth with very low dislocation density. In the second part we have discussed the doping level optimization in the GaN device stack. On engineered substrates a net doping level of $1.5\times10^{16}$ cm$^{-3}$ has been achieved in the n-drift layer. A reference device using an 800 nm p-GaN channel with a $1.2\times10^{19}$ Mg/cm$^3$ doping has been demonstrated, showing a current density of 0.35 kA/cm$^2$ and a reasonably low R$_{ON}$ of 10 mΩ·cm$^2$. In the last part the trench gate reliability has been discussed, for which it is essential to implement a process which results in a smooth trench gate sidewall, combined with the use of a sufficiently thick gate dielectric.

**Acknowledgements**

This project has received funding from the ECSEL Joint Undertaking (JU) under grant agreement No 826392. The JU receives support from the European Union's Horizon 2020 research and innovation program and Austria, Belgium, Germany, Italy, Slovakia, Spain, Sweden, Norway, Switzerland.

**References**

[1] J. Hu et al., "Materials and processing issues in vertical GaN power electronics", Materials Science in Semiconductor Processing, Volume 78, 2018, Pages 75-84, https://doi.org/10.1016/j.mssp.2017.09.033.

[2] Y. Sun et al., "Review of the Recent Progress on GaN-Based Vertical Power Schottky Barrier Diodes (SBDs)", *Electronics* 2019, *8*,575. https://doi.org/10.3390/electronics8050575

[3] NexGen Power Systems's vertical GaN$^{TM}$ Technology—the power conversion platform of choice, https://nexgenpowersystems.com/verticalgan/, May 2021.

[4] D. Shibata, et al., "1.7 kV/1.0 mΩcm$^2$ normally-off vertical GaN transistor on GaN substrate with regrown p-GaN/AlGaN/GaN semipolar gate structure," *2016 IEEE International Electron Devices Meeting (IEDM)*, 2016, pp. 10.1.1-10.1.4. doi: 10.1109/IEDM.2016.7838385.

[5] N. M. Shrestha, "A novel AlGaN/GaN multiple aperture vertical high electron mobility transistor with silicon oxide current blocking layer", Vacuum, Volume 118, 2015, Pages 59-63, https://doi.org/10.1016/j.vacuum.2014.11.022.

[6] Y. Zhang et al., "1200V GaN vertical Fin power field-effect transistors", 2017 IEEE International Electron Devices Meeting (IEDM), 2017, pp. 9.2.1-9.2.4, doi: 10.1109/IEDM.2017.8268357.

[7] H. Ohta et al., "Vertical GaN p-n Junction Diodes with High Breakdown Voltages Over 4 kV," in IEEE Electron Device Letters, vol. 36, no. 11, pp. 1180-1182, Nov. 2015, doi: 10.1109/LED.2015.2478907.

[8] W. Li et al., "Development of GaN Vertical Trench-MOSFET With MBE Regrown Channel," in IEEE Transactions on Electron Devices, vol. 65, no. 6, pp. 2558-2564, June 2018, doi: 10.1109/TED.2018.2829125.

[9] I. C. Kizilyalli et al., "Vertical Power p-n Diodes Based on Bulk GaN," in IEEE Transactions on Electron Devices, vol. 62, no. 2, pp. 414-422, Feb. 2015, doi: 10.1109/TED.2014.2360861.

[10] H. Ohta et al., "Impact of Lowering Threading Dislocation Density on Performances of Vertical GaN p-n Junction Diodes," *2019 Compound Semiconductor Week (CSW)*, 2019, pp. 1-2, doi: 10.1109/ICIPRM.2019.8819005.

[11] X. Zou et al., "Breakdown Ruggedness of Quasi-Vertical GaN-Based p-i-n Diodes on Si Substrates," in *IEEE Electron Device Letters*, vol. 37, no. 9, pp. 1158-1161, Sept. 2016, doi: 10.1109/LED.2016.2594821.

[12] Y. Zhang et al., "720-V/0.35-mΩ·cm$^2$ Fully Vertical GaN-on-Si Power Diodes by Selective Removal of Si Substrates and Buffer Layers," in *IEEE Electron Device Letters*, vol. 39, no. 5, pp. 715-718, May 2018, doi: 10.1109/LED.2018.2819642.

[13] Y. Zhang *et al*., "High-Performance 500 V Quasi- and Fully-Vertical GaN-on-Si pn Diodes," in *IEEE Electron Device Letters*, vol. 38, no. 2, pp. 248-251, Feb. 2017, doi: 10.1109/LED.2016.2646669.

[14] Y. Zhang *et al*., "Novel GaN trench MIS barrier Schottky rectifiers with implanted field rings," *2016 IEEE International Electron Devices Meeting (IEDM)*, 2016, pp. 10.2.1-10.2.4, doi: 10.1109/IEDM.2016.7838386.

[15] X. Zhang et al., "Fully- and Quasi-Vertical GaN-on-Si p-i-n Diodes: High Performance and Comprehensive Comparison," in *IEEE Transactions on Electron Devices*, vol. 64, no. 3, pp. 809-815, March 2017, doi: 10.1109/TED.2017.2647990.

[16] R. Abdul Khadar et al., "820-V GaN-on-Si Quasi-Vertical p-i-n Diodes with BFOM of 2.0 GW/cm$^2$,"in *IEEE Electron Device Letters*, vol. 39, no. 3, pp. 401-404, March 2018, doi: 10.1109/LED.2018.2793669.

[17] R. A. Khadar et al., "Fully Vertical GaN-on-Si power MOSFETs," in IEEE Electron Device Letters, vol. 40, no. 3, pp. 443-446, March 2019, doi: 10.1109/LED.2019.2894177.

[18] V. Odnoblyudov et al., U.S. Patent, US10297445B2, May 21, 2019.
[19] T.J. Anderson et al., "Electrothermal evaluation of thick GaN epitaxial layers and AlGaN/GaN high-electron-mobility transistors on large-area engineered substrates". Appl. Phys. Express, vol. 10, no. 12, pp. 126501-1–126501-3, Nov. 2017, doi: https://doi.org/10.7567/APEX.10.126501.J. Clerk Maxwell, A Treatise on Electricity and Magnetism, 3rd ed., vol. 2. Oxford: Clarendon, 1892, pp.68–73.
[20] A. Zubair et al., "First demonstration of GaN vertical power FinFETs on engineered substrate," 2020 Device Research Conference (DRC), 2020, pp. 1-2. doi: 10.1109/DRC50226.2020.9135176.
[21] A. D. Koehler et al., "Vertical GaN SBDs on large engineered substrates", Workshop on compound semiconductor materials & devices (WOCSEMMAD) 2017, February 19-22, Safety Harbor, Florida, USA, 2017.
[22] T. J. Anderson et al., "Lateral GaN JFET Devices on 200 mm Engineered Substrates for Power Switching Applications," 2018 IEEE 6th Workshop on Wide Bandgap Power Devices and Applications (WiPDA), 2018, pp. 14-17, doi: 10.1109/WiPDA.2018.8569101
[23] T. Hikosaka et al., "Growth of high-quality > 10 μm-thick GaN-on-Si with low-dislocation density in the order of $10^7/cm^2$", Compound Semiconductor Week 2019, May 19 to 23, 2019, Nara, Japan. doi: 10.1109/ICIPRM.2019.8819020.
[24] A. Tanaka et al., "Structural and electrical characterization of thick GaN layers on Si, GaN and engineered substrates", Journal of Applied Physics 125, 082517 (2019). https://doi.org/10.1063/1.5049393.
[25] K. Geens et al., "650 V p-GaN Gate Power HEMTs on 200 mm Engineered Substrates," in IEEE 7th Workshop on Wide Bandgap Power Devices and Applications (WiPDA), Raleigh, NC, USA, 2019, pp. 292-296, doi: 10.1109/WiPDA46397.2019.8998922.
[26] X. Li et al., "Integration of 650 V GaN Power ICs on 200 mm Engineered Substrates," in IEEE Transactions on Semiconductor Manufacturing, vol. 33, no. 4, pp. 534-538, Nov. 2020, doi: 10.1109/TSM.2020.3017703.
[27] X. Li et al., "Demonstration of GaN Integrated Half-Bridge with On-Chip Drivers on 200-mm Engineered Substrates," in IEEE Electron Device Letters, vol. 40, no. 9, pp. 1499-1502, Sept. 2019, doi: 10.1109/LED.2019.2929417.
[28] K. Geens et al., "Demonstration of high-quality GaN epitaxy on 200mm engineered substrates for vertical power device fabrication", CS Mantech, 24-27 May, 2021.
[29] Shin-Etsu Chemical shifts into high gear in the development of gallium nitride substrates and related products. https://www.shinetsu.co.jp/en/news/shin-etsu-chemical-shifts-into-high-gear-in-the-development-of-gallium-nitride-substrates-and-related-products/. (May 2021)
[30] GaN on QST[®] templates, https://www.kymatech.com/products-services/materials/gan-related-iii-n-materials/500-200-mm-hvpe-gan-on-qst-templates, (May 2021).
[31] M. A. Moram et al., "X-Ray diffraction of III-nitrides", in reports on progress in Physics, vol. 72, no. 3, Feb. 2009.
[32] J. Colvin et al., "Surface and dislocation investigation of planar GaN formed by crystal reformation of nanowire arrays", Phys. Rev. Materials, Vol. 3, No. 9, pp. 093604, September 2019, American Physical Society, doi: 10.1103/PhysRevMaterials.3.093604.
[33] M. Khalilian et al., "Dislocation-Free and Atomically Flat GaN Hexagonal Microprisms for Device Applications", Small 2020, 16, 907364. https://doi.org/10.1002/smll.201907364
[34] D. Carrascon et al., "Optimization of GaN Nanowires Reformation Process by Metalorganic Chemical Vapor Deposition for Device-Quality GaN Templates", Phys. Status Solidi B, 257: 1900581. https://doi.org/10.1002/pssb.201900581
[35] B J Baliga, "Power semiconductor device figure of merit for high-frequency applications", IEEE Electron Device Letters. 1989;10(10):455–457.
[36] T. Ciarkowski et al., "Connection between Carbon Incorporation and Growth Rate for GaN Epitaxial Layers Prepared by OMVPE", Materials 2019, 12, 2455. https://doi.org/10.3390/ma12152455
[37] A. Agarwal et al., "Controlled low Si doping and high breakdown voltages in GaN on sapphire grown by MOCVD", Semicond. Sci. Technol., Bd. 31, Nr. 125018, 2016.
[38] R. Hentschel et al., "Extraction of the active acceptor concentration in (pseudo-) vertical GaN MOSFETs using the body-bias effect", Microelectron. J., Bd. 91, pp. 42-45, 2019.
[39] K. Mukherjee et al., "Understanding the Leakage Mechanisms and Breakdown Limits of Vertical GaN-on-Si p[+]n[−]n Diodes: The Road to Reliable Vertical MOSFETs", Micromachines 2021, 12, 445. https://doi.org/10.3390/mi12040445
[40] K. Mukherjee, et al., "Challenges and Perspectives for Vertical GaN-on-Si Trench MOS Reliability: From Leakage Current Analysis to Gate Stack Optimization", Materials 2021, 14, 2316. https://doi.org/10.3390/ma14092316
[41] K. Mukherjee et al., "Use of Bilayer Gate Insulator in GaN-on-Si Vertical Trench MOSFETs: Impact on Performance and Reliability", Materials 2020, 13, 4740. https://doi.org/10.3390/ma13214740
[42] P. Diehle et al., "Root cause analysis of gate shorts in semi-vertical GaN MOSFET devices," in The 13th International Conference on Advanced Semiconductor Devices And Microsystems (ASDAM), Smolenice, Slovakia, November 2020, pp. 10-13.